\documentclass[11pt]{article}
\usepackage[T1]{fontenc}
\usepackage{lmodern,microtype,mathtools,amssymb,amsthm,booktabs,array}
\usepackage[margin=1in]{geometry}
\usepackage[colorlinks=true,linkcolor=blue,citecolor=blue,urlcolor=blue]{hyperref}
\hypersetup{pdftitle={Maximal input entropy for binary two-deletion and two-insertion channels},pdfauthor={Haobo Ma and Wenlin Zhang},pdflang={en}}
\newtheorem{theorem}{Theorem}[section]
\newtheorem{lemma}[theorem]{Lemma}
\newtheorem{proposition}[theorem]{Proposition}
\newtheorem{corollary}[theorem]{Corollary}
\newtheorem{conjecture}[theorem]{Conjecture}
\theoremstyle{remark}
\newcommand{\bits}{\{0,1\}}
\newcommand{\DD}{\Phi_{D,2}}
\newcommand{\II}{\Phi_{I,2}}
\newcommand{\HD}{H_{D,k}}
\newcommand{\HI}{H_{I,k}}
\DeclareMathOperator*{\argmax}{arg\,max}
\title{Maximal input entropy for binary two-deletion and two-insertion channels}
\author{Haobo Ma$^1$\thanks{ORCID: \href{https://orcid.org/0009-0008-4944-977X}{0009-0008-4944-977X}.}
\and Wenlin Zhang$^{1,2}$\thanks{Corresponding author: \href{mailto:e1327962@u.nus.edu}{e1327962@u.nus.edu}.
ORCID: \href{https://orcid.org/0009-0005-2414-7968}{0009-0005-2414-7968}.}}
\date{31 October 2026}
\begin{document}
\maketitle
\begin{center}
$^1$ChronoAI Pte Ltd, Singapore\\
$^2$National University of Singapore, Singapore
\end{center}
\begin{abstract}
Under uniform transmission, we determine the received binary words that maximize input entropy for two-deletion and two-insertion channels. For two deletions the maximizers are exactly the two alternating words at every positive received length. For two insertions the same conclusion holds from received length five onward; we give all smaller cases and exact maximal values. The proof combines an embedding-count identity organized by runs with strict convexity estimates, and a grouping of two-position deletion masks. For every insertion radius we give the complete positive embedding histogram of an alternating received word. We also determine its three-deletion histogram for every length, and certify all three-deletion maximizers through received length twelve and all three-insertion maximizers through length eighteen by exact integer comparison. These computations give precise three-error conjectures with their small-length exceptions.
\end{abstract}

\section{Introduction}
We determine which received binary words maximize the posterior input entropy for exactly two deletions or exactly two insertions. The posterior uncertainty depends on subsequence embedding multiplicities. Even with a uniform input distribution, the posterior is generally far from uniform: different index sets can produce the same word. Extremizing this conditional entropy therefore requires the full distribution of these multiplicities.

The posterior-entropy problem was investigated by Atashpendar, Roscoe and Ryan~\cite{Atashpendar2015}, who observed constant and alternating extremizers in their study of information leakage. Atashpendar et al.~\cite{Atashpendar2019} subsequently studied the posterior distribution by clustering supersequences and using run lengths, and describe the conjecture that constant words minimize the entropy and alternating words maximize it. The related asymptotic work~\cite{Atashpendar2018} proves entropy minimization when the received length is fixed and the transmitted length tends to infinity. Following their ITW paper~\cite{Singhvi2022}, Singhvi, Sabary, Bar-Lev and Yaakobi~\cite{Singhvi2025} give embedding formulas for both channels, settle the one-error extrema, and determine the two-deletion minimum and the general insertion minimum. In the latest arXiv version, Section VII-A, first unnumbered bullet, they state:
\begin{quote}
``\textbf{Maximal Input Entropy for $k=2$.} While we provide a complete characterization of the sequences that minimize input entropy, the precise identification of the sequences that \emph{maximize} input entropy for the $2$-deletion and $2$-insertion channels is still unresolved. Determining these sequences and the associated maximal entropy values remains an important open direction.''
\end{quote}
This is arXiv:2407.10026v2, p.~16, of the work published as~\cite{Singhvi2025}. Their next bullet asks for extensions to higher radii.

We answer the two-error question completely. Alternating received words uniquely maximize the two-deletion input entropy. For two insertions they uniquely maximize it from length five onward, with an exceptional maximizing pair at length four. The deletion proof rearranges the run and alternating-segment enumeration of~\cite[Theorem 3]{Singhvi2025} into a form suited to a strict lower bound for the entropy correction. We give a self-contained derivation by adding the first symbol. For insertion, grouping masks gives a lower bound sufficient for a comparison even without classifying every possible coincidence between subsequences.

Embedding histograms also arise in the enumeration of words with a prescribed number of occurrences of a fixed subsequence. Menon and Singh~\cite{MenonSingh2024} obtain host-word counts for multiplicities one through four; their formulas recover four rows of our three-deletion histogram. Bruda~\cite{Bruda2026} proves eventual polynomial enumeration for a fixed nonconstant pattern and fixed positive occurrence count, and gives a run-based embedding bound sharp when both words are alternating.

For higher radii we establish complete alternating-word histograms. The positive embedding histogram of an alternating received word has a binomial expression at every insertion radius. At radius three we also enumerate its supersequence multiplicities in closed form. Exact exhaustive comparisons support alternating maximizers for three deletions at all lengths checked, and for three insertions from length seven onward. The all-length optimality assertions at radius three remain conjectures.

Throughout the paper, $n$ is the \emph{received} length. Thus the transmitted length is $n+k$ for $k$ deletions and $n-k$ for $k$ insertions. In~\cite{Singhvi2025}, $n$ instead denotes transmitted length. All logarithms below have base two.

\section{Embedding multiplicities and the main theorem}
For finite binary words $u$ and $v$, let $e(u,v)$ be the number of increasing index sets of $v$ whose selected symbols form $u$. Set $e(\varnothing,v)=1$. Let $A_n=0101\cdots$ be the alternating word of length $n$, and write $\overline{y}$ for bitwise complementation. Define
\[
 f(t)=t\log t\quad(t>0),\qquad f(0)=0,
\]
and
\begin{align}
 \Phi_{D,k}(y)&=\sum_{x\in\bits^{n+k}}f(e(y,x)),&
 Z_{D,k}&=2^k\binom{n+k}{k},\label{eq:DPhi}\\
 \Phi_{I,k}(y)&=\sum_{x\in\bits^{n-k}}f(e(x,y)),&
 Z_{I,k}&=\binom nk.\label{eq:IPhi}
\end{align}
For insertion we assume $n\ge k$. The channel chooses an error-position set uniformly, and insertion symbols are independent uniform bits. Uniform transmission means that every input word of the prescribed length has the same prior probability.

\begin{lemma}\label{lem:entropy}
The posterior probabilities of input words, given $y$, are $e(y,x)/Z_{D,k}$ and $e(x,y)/Z_{I,k}$, respectively. Consequently
\begin{equation}\label{eq:entropy}
 \HD(y)=\log Z_{D,k}-\frac{\Phi_{D,k}(y)}{Z_{D,k}},\qquad
 \HI(y)=\log Z_{I,k}-\frac{\Phi_{I,k}(y)}{Z_{I,k}}.
\end{equation}
\end{lemma}
\begin{proof}
For deletion, each embedding of $y$ in $x$ specifies the retained positions, so the likelihood is $e(y,x)/\binom{n+k}{k}$. Sum $e(y,x)$ over $x$ by choosing the $k$ extra positions and their bits; this gives $Z_{D,k}$. The uniform prior and Bayes' rule give the first posterior. For insertion, each embedding of $x$ in $y$ specifies the inserted positions and their required bits. The likelihood is $e(x,y)/(2^k\binom nk)$. Summing the embedding counts over $x$ counts all $\binom nk$ deletion masks on $y$, giving the second posterior. Substitute these posteriors into Shannon entropy.
\end{proof}
These are the binary specializations of~\cite[Lemmas 2 and 3]{Singhvi2025}.

\begin{theorem}\label{thm:two}
For every $n\ge1$,
\begin{equation}\label{eq:Dmax}
\begin{gathered}
 \argmax_{y\in\bits^n}H_{D,2}(y)=\{A_n,\overline{A_n}\},\\
 \max_y H_{D,2}(y)=\log\!\left(4\binom{n+2}{2}\right)
 -\frac{4n^2+3n\log3+(n+1)\log(n+1)}{4\binom{n+2}{2}}.
\end{gathered}
\end{equation}
For every $n\ge5$,
\begin{equation}\label{eq:Imax}
\begin{gathered}
 \argmax_{y\in\bits^n}H_{I,2}(y)=\{A_n,\overline{A_n}\},\\
 \max_y H_{I,2}(y)=\log\binom n2-\frac{(n-1)\log(n-1)}{\binom n2}.
\end{gathered}
\end{equation}
At received length two every word maximizes, with value zero; at length three exactly the six nonconstant words maximize; at length four exactly $0110$ and $1001$ maximize.
\end{theorem}
The maximal insertion entropies at received lengths three and four are, respectively, $\log3-2/3$ and $\log6-2/3$, as computed in Section~4.2.

\section{The two-deletion comparison}
\subsection{A run identity}
Write the positive run lengths of $y$ as $r_1,\ldots,r_R$. List those at least two, in their original order, as $u_1,\ldots,u_q$. Put $U=\sum_j u_j$, and let $s_0,\ldots,s_q$ count the singleton runs before, between and after these runs. If $q>0$, define
\begin{equation}\label{eq:bridges}
 w_0=u_1+s_0+1,\quad w_q=u_q+s_q+1,\quad
 w_j=u_j+u_{j+1}+s_j+1\quad(1\le j<q).
\end{equation}
Also put
\[
 B_n=f(3)+4n,\qquad h_n(r)=f\!\left(\binom{r+2}{2}\right)+(2n-r)f(r+1).
\]

\begin{lemma}\label{lem:run}
For $q>0$,
\begin{equation}\label{eq:run}
 \DD(y)=(n-U)B_n+\sum_{j=1}^q h_n(u_j)+\sum_{j=0}^q f(w_j).
\end{equation}
For $q=0$ the value is $nB_n+f(n+1)$.
\end{lemma}
\begin{proof}
We derive the identity through an exact first-symbol recurrence. Fix a tail run list $T=(t_1,\ldots,t_\ell)$ of total length $s$. Let $L(T)$ be the number of its initial singleton runs plus the length of its first nonsingleton run; set $L(T)=s$ if all runs are singleton, including $L(\varnothing)=0$. Let $D(r,T)$ denote $\DD$ for a word with first run of length $r$ followed by $T$; at $r=0$ the first run is removed. Complementation makes the choice of first bit immaterial. We claim
\begin{align}
 D(r+1,T)-D(r,T)
 ={}&f\!\left(\binom{r+3}{2}\right)-f\!\left(\binom{r+2}{2}\right)\notag\\
 &+(r+2+2s)f(r+2)-(r+1+2s)f(r+1)\notag\\
 &+2\sum_{i=1}^{\ell}f(t_i+1)
 +f(r+2+L(T))-f(r+1+L(T))\label{eq:increment}
\end{align}
for every integer $r\ge0$.

Write $\operatorname{runWord}(c,T)$ for the binary word with first bit $c$ and successive run lengths $T$, taking the empty word when $T=\varnothing$. To see the recurrence, let $w=b^{r+1}\operatorname{runWord}(\bar b,T)$ and let $v$ be $w$ with its first symbol removed. For words $z$ of length $|w|+1$, put $a=e(w,z)$ and $d=e(v,z)$. Those with $a>0$ have the following joint counts:
\begin{center}
\begin{tabular}{lll}
\toprule
$a$&$d$&number of distinct words $z$\\
\midrule
$r+2$&$\binom{r+2}{2}$&$1$\\
$t_i+1$&$(r+1)(t_i+1)$&$1$ for each tail run $i$\\
$1$&$r+1$&$r+1+s-\ell$\\
$1$&$r+1+L(T)$&$1$\\
\bottomrule
\end{tabular}
\end{center}
For completeness, every one-symbol supersequence is obtained either by extending a run (the first two rows), or by inserting the other symbol inside a run or at an end. The latter placements yield $|w|+2-(\ell+1)=r+s+2-\ell$ distinct words with $a=1$. In all but one of them, deleting the leading $b$ leaves $r+1$ embeddings of $v$. The exceptional word is $b^r\bar b b\operatorname{runWord}(\bar b,T)$, obtained by inserting $\bar b$ immediately before the final $b$ of the first run. In this word, the position omitted before the tail can move successively through its initial singleton runs and into its first nonsingleton run, giving $L(T)$ additional embeddings. This also describes the empty-tail case, where the last two rows happen to have equal weights. Thus the table covers every distinct one-symbol supersequence once.

Separating the leading bit of a two-symbol supersequence gives
\[
 e(w,bz)=a+d,\qquad e(w,\bar b z)=a.
\]
Hence $D(r+1,T)-D(r,T)=\sum_z\{f(a+d)-f(d)+f(a)\}$. Terms with $a=0$ vanish. The sum of $f(a)$ is $f(r+2)+\sum_i f(t_i+1)$. Substitution of the table and $f(ab)=af(b)+bf(a)$ gives~\eqref{eq:increment}.

Now let $F$ denote the proposed right side of~\eqref{eq:run}, with the stated all-singleton convention and $F(\varnothing)=0$. We verify the same increment. For $r\ge2$, the singleton contribution changes by four per singleton tail run; each nonsingleton tail run contributes $2f(t_i+1)$. The first-run term changes by
\[
 f\!\left(\binom{r+3}{2}\right)-f\!\left(\binom{r+2}{2}\right)
 +(r+1+2s)f(r+2)-(r+2s)f(r+1).
\]
The two affected weights in~\eqref{eq:bridges} contribute
\[
 f(r+2)-f(r+1)+f(r+2+L(T))-f(r+1+L(T)),
\]
which gives~\eqref{eq:increment}. At $r=0$, adding a singleton contributes $B_{s+1}$, four per singleton tail run, $2f(t_i+1)$ per nonsingleton tail run, and $f(L(T)+2)-f(L(T)+1)$. At $r=1$, turning the first singleton into a run of length two contributes, apart from the tail increments,
\[
 f(6)+2(s+1)f(3)-4(s+1)+f(L(T)+3)-f(L(T)+2).
\]
These are exactly the $r=0$ and $r=1$ instances of~\eqref{eq:increment}. Induction on word length, starting from the empty word, proves the identity.
\end{proof}

\subsection{Strict minimization of the correction}
Put $a=\log3$. The exact inequalities $2^3<3^2$ and $3^3<2^5$ give
\begin{equation}\label{eq:a}
 \frac32<a<\frac53.
\end{equation}
Let $c_n=(6a-8)n+6-6a$. We first show
\begin{equation}\label{eq:hn}
 h_n(r)\ge rB_n+c_n\qquad(2\le r\le n).
\end{equation}
At $r=2$ this is equality. For $r\ge3$, convexity gives $f(r+1)-f(r)\ge f(3)-f(2)=3a-2$. It also gives
\[
 f(v+w)-f(v)\ge f(w)+2w\qquad(v\ge w>0),
\]
because the increment is increasing in $v$ and its value at $v=w$ is $f(2w)-f(w)=f(w)+2w$. Apply this with $v=r(r+1)/2$ and $w=r+1$. Expanding the difference of the $h_n$ terms yields
\begin{align*}
 h_n(r)-h_n(r-1)-B_n
 &\ge (6a-8)n+(4-3a)r\\
 &\ge (3a-4)n>0.
\end{align*}
The second inequality uses $r\le n$ and $4-3a<0$. This proves~\eqref{eq:hn} by induction.

Suppose now $y$ is not alternating, so $q\ge1$ and $U\ge2q$. By~\eqref{eq:bridges},
\[
 \sum_{j=0}^q w_j=n+U+q+1\ge S:=n+3q+1.
\]
All weights are positive, and $t\log(t/(q+1))$ is increasing for $t\ge q+1$. Jensen's inequality and~\eqref{eq:hn} give
\begin{equation}\label{eq:JensenD}
 \DD(y)\ge nB_n+qc_n+S\log\frac{S}{q+1}.
\end{equation}
Since $S>n+1$, $S/(q+1)\ge3$, and $\log(q+1)\le q$,
\begin{align*}
 S\log\frac{S}{q+1}
 &=(n+1)\log\frac{S}{q+1}+3q\log\frac{S}{q+1}\\
 &>f(n+1)-(n+1)q+3qa.
\end{align*}
Thus the right side of~\eqref{eq:JensenD} exceeds $nB_n+f(n+1)$ by at least
\[
 q\bigl((6a-9)n+5-3a\bigr)>0,
\]
using~\eqref{eq:a}. Lemma~\ref{lem:run} gives $\DD(A_n)=4n^2+3n\log3+f(n+1)$. Therefore only $A_n$ and its complement minimize $\DD$. Lemma~\ref{lem:entropy} proves~\eqref{eq:Dmax}.

\section{The two-insertion comparison}
\subsection{Grouping two-position masks}
The same-run, nonadjacent-run and alternating-segment cases are enumerated in~\cite[Theorem 4]{Singhvi2025}. Here we group adjacent masks into components to obtain a lower bound sufficient for the strict extremal estimates below.
For run lengths $r_1,\ldots,r_R$, a mask removing two positions in run $i$ has total weight $\binom{r_i}{2}$, and masks removing one position in each of distinct runs $i,j$ have total weight $r_i r_j$. For adjacent pairs begin with edges of weights $r_i r_{i+1}$. Join successive edges exactly when their shared run is singleton, and denote the component sums by $v_1,\ldots,v_c$. Deleting that shared singleton allows the removed position in the neighboring run to slide from one edge to the next, so every component produces a single subsequence. Since
\[
 f(a)+f(b)\le f(a+b)\qquad(a,b\ge0),
\]
any further coincidences between the resulting subsequences can only increase their contribution. Consequently
\begin{equation}\label{eq:P}
 P(y):=\sum_i f\!\left(\binom{r_i}{2}\right)
 +\sum_{i<j-1}f(r_i r_j)+\sum_{\ell=1}^c f(v_\ell)
 \le\II(y).
\end{equation}
For an alternating word all adjacent edges belong to one component of weight $n-1$. Every nonadjacent mask gives a different subsequence of multiplicity one: its two removed positions are recovered from the two resulting equal-symbol adjacencies, or from the affected endpoint. Equivalently the $k=2$ case of Theorem~\ref{thm:allI} below gives the same histogram. Hence
\begin{equation}\label{eq:altI2}
 \II(A_n)=f(n-1)\qquad(n\ge3).
\end{equation}

Let $d=n-R=\sum_i(r_i-1)$. We prove two complementary bounds:
\begin{align}
 P(y)&\ge2(n-2)d,\label{eq:bound1}\\
 P(y)&\ge f(n-1)+d(n-2d-5)\quad\hbox{if }R\ge3.\label{eq:bound2}
\end{align}
For~\eqref{eq:bound1}, first unmerge adjacent edges in~\eqref{eq:P}. Since $f(ab)=af(b)+bf(a)$, the resulting sum is
\[
 \sum_i\left[f\!\left(\binom{r_i}{2}\right)+(n-r_i)f(r_i)\right].
\]
For every integer $r\ge1$, convexity implies $f(r)\ge2(r-1)$ and
\begin{equation}\label{eq:binomineq}
 f\!\left(\binom r2\right)\ge2(r-1)(r-2).
\end{equation}
The latter is direct for $r=1,2,3$. For $r\ge4$ apply $f(t)\ge2(t-1)$, valid for $t\ge2$, to $t=r/2$ and $t=r-1$, and expand $f((r/2)(r-1))$. The resulting lower bound is $2r^2-5r+2\ge2(r-1)(r-2)$. These inequalities give~\eqref{eq:bound1}.

For~\eqref{eq:bound2}, the nonadjacent term satisfies
\[
 \sum_{i<j-1}f(r_i r_j)
 =\sum_i f(r_i)\!\sum_{j:\,|i-j|\ge2}r_j
 \ge(R-3)\sum_i f(r_i)\ge2d(R-3).
\]
Also $\sum_\ell v_\ell=\sum_i r_i r_{i+1}\ge n-1$: sum $ab\ge a+b-1$ over edges. Only an internal nonsingleton run separates edge components, so $c\le d+1$. For $R\ge3$ we have $d+1\le n-1$, and Jensen yields
\[
 \sum_\ell f(v_\ell)\ge(n-1)\log\frac{n-1}{d+1}
 \ge f(n-1)-(n-1)d.
\]
Here the monotonicity of $t\log(t/c)$ for $t\ge c$ justifies replacing the total edge weight by $n-1$. Drop the nonnegative same-run term and substitute $R=n-d$ to obtain~\eqref{eq:bound2}.

\subsection{The comparison and the finite initial cases}
Suppose $n\ge9$ and $y$ is not alternating. Then $d\ge1$. If $2d<n-5$, necessarily $R\ge3$, and~\eqref{eq:bound2} is strictly larger than $f(n-1)$. Otherwise~\eqref{eq:bound1} gives
\[
 P(y)\ge(n-2)(n-5)>f(n-1).
\]
To verify the last inequality, $\log(n-1)\le(n-3)/2$ for integers $n\ge9$. Indeed, $t^2\le2^{t-2}$ at $t=8$, and induction follows from $(1+1/t)^2<2$ for $t\ge8$. Finally
\[
 (n-2)(n-5)-\frac{(n-1)(n-3)}2
 =\frac{n^2-10n+17}{2}>0\quad(n\ge9).
\]
Together with~\eqref{eq:P} and~\eqref{eq:altI2}, this proves the strict comparison for $n\ge9$.

The remaining cases are finite exact comparisons. For any channel with integer embedding counts, minimizing the correction is equivalent to minimizing
\begin{equation}\label{eq:Q}
 Q(y)=\prod_{x:e>0}e^{\,e},
\end{equation}
where $e=e(y,x)$ for deletion and $e=e(x,y)$ for insertion. In fact $\log Q(y)=\Phi(y)$. The following table results from enumerating the $2^n$ words and all $\binom n2$ masks on each word.
\begin{center}
\begin{tabular}{clrl}
\toprule
$n$&minimizing words&$\min Q$&number\\
\midrule
2&all words&1&4\\
3&all nonconstant words&4&6\\
4&$0110,1001$&16&2\\
5&$A_5,\overline{A_5}$&256&2\\
6&$A_6,\overline{A_6}$&3125&2\\
7&$A_7,\overline{A_7}$&46656&2\\
8&$A_8,\overline{A_8}$&823543&2\\
\bottomrule
\end{tabular}
\end{center}
The exact finite algorithm and its completeness justification are in Section~\ref{sec:certificates}. In particular, the two cases $n=3,4$ have $\Phi=2,4$, respectively. At $n=4$ the counts for the possible length-two inputs $00,01,10,11$ are $(1,2,2,1)$ for $0110$, versus $(1,3,1,1)$ for $0101$. The advantage is $(3\log3-4)/6>0$. The table, the strict comparison and Lemma~\ref{lem:entropy} finish the proof of Theorem~\ref{thm:two}.

\section{Higher radii: exact alternating-word formulas}
We use the convention $\binom ab=0$ when $b<0$ or $b>a$, for nonnegative integer $a$. A positive embedding histogram lists each multiplicity together with the number of distinct words attaining it. If displayed multiplicities coincide, their counts are added.

\subsection{All insertion radii}
\begin{theorem}\label{thm:allI}
Let $m\ge1$, $k\ge0$, and $n=m+k$. The positive embedding histogram of $e(x,A_n)$ over $x\in\bits^m$ is
\begin{equation}\label{eq:allI}
 \boxed{\quad
 e_t=\binom{m+t}{t},\qquad N_t=\binom{m+1}{k-2t},
 \quad 0\le t\le\lfloor k/2\rfloor .\quad}
\end{equation}
Rows with $N_t=0$ are omitted. In particular,
\begin{equation}\label{eq:allIentropy}
 H_{I,k}(A_n)=\log\binom nk
 -\frac{\displaystyle\sum_{t=0}^{\lfloor k/2\rfloor}
 \binom{m+1}{k-2t} f\!\left(\binom{m+t}{t}\right)}{\binom nk}.
\end{equation}
When $m=0$, there is one possible input, the empty word, and the entropy is zero.
\end{theorem}
\begin{proof}
Fix $x=x_1\cdots x_m$, write $b=x_1\in\{0,1\}$, and let $d$ be the number of equal adjacent pairs in $x$. The earliest possible first embedding position in $A_n$ is $1+b$. The earliest increment between consecutive selected positions is one if the corresponding symbols differ, and two if they agree. Thus the earliest last position is $m+d+b$.

All other embeddings are obtained uniquely by adding an even nonnegative increment $2a_0$ to the first position and $2a_j$ to the $j$th gap, $1\le j<m$. With $t=\lfloor(k-d-b)/2\rfloor$, the requirement is $\sum_{j=0}^{m-1}a_j\le t$. There are no embeddings if $t<0$; otherwise stars and bars gives $\binom{m+t}{m}$ embeddings. For each initial bit $b$, there are $\binom{m-1}{d}$ choices of $x$ with $d$ equal adjacent pairs: specify exactly which adjacent pairs are equal. Therefore the number of words with $d+b=s$ is
\[
 \binom{m-1}{s}+\binom{m-1}{s-1}=\binom ms.
\]
The words producing a fixed nonnegative $t$ have $s=k-2t$ or $s=k-2t-1$. Their total number is $\binom{m+1}{k-2t}$, proving~\eqref{eq:allI}. Equation~\eqref{eq:allIentropy} follows from Lemma~\ref{lem:entropy}.
\end{proof}
As a consistency identity, summing multiplicity times count gives
\[
 \sum_{t=0}^{\lfloor k/2\rfloor}\binom{m+1}{k-2t}\binom{m+t}{t}=\binom{m+k}{k}.
\]
This also follows from the generating function $(1+z)^{m+1}(1-z^2)^{-m-1}=(1-z)^{-m-1}$.

\begin{corollary}\label{cor:I3}
For $n\ge4$, the three-insertion histogram of $A_n$ has multiplicity $n-2$ with count $n-2$, and multiplicity one with count $\binom{n-2}{3}$. Hence
\begin{equation}\label{eq:I3}
 H_{I,3}(A_n)=\log\binom n3-\frac{(n-2)^2\log(n-2)}{\binom n3}.
\end{equation}
\end{corollary}
\begin{proof}
In~\eqref{eq:allI} take $m=n-3$. The rows $t=1$ and $t=0$ give the two stated multiplicities and counts.
\end{proof}
For example, radius four has one word of multiplicity $\binom{m+2}{2}$, $\binom{m+1}{2}$ words of multiplicity $m+1$, and $\binom{m+1}{4}$ words of multiplicity one. The formulas describe the alternating word's entropy; optimality is a separate comparison problem.

\subsection{Three deletions}
For $n\ge4$, the multiplicity-one, -two, -three and -four counts below also follow from~\cite[Propositions 3.2, 3.4, 3.8 and 3.11]{MenonSingh2024}: take the pattern $A_n$, with $n$ singleton runs, and host length $n+3$. The four counts specialize to $4$, $3n$, $2n$ and $n+2\binom n2=n^2$. We give a direct proof of the complete histogram, including all smaller lengths.
\begin{theorem}\label{thm:D3}
For every $n\ge1$, the positive embedding histogram of $e(A_n,x)$ over $x\in\bits^{n+3}$ is the following table, with coincident rows combined:
\begin{center}
\begin{tabular}{cr@{\qquad}cr@{\qquad}cr}
\toprule
multiplicity&count&multiplicity&count&multiplicity&count\\
\midrule
1&4&2&$3n$&3&$2n$\\
4&$n^2$&6&$n(n-1)$&8&$\binom n3$\\
$2n$&$n$&$2n+1$&2&$n+1$&2\\
\bottomrule
\end{tabular}
\end{center}
Thus, with
\begin{align}\label{eq:SD3}
 S_3(n)={}&3n f(2)+2n f(3)+n^2f(4)+n(n-1)f(6)+\binom n3 f(8)\notag\\
 &+n f(2n)+2f(2n+1)+2f(n+1),
\end{align}
we have
\begin{equation}\label{eq:D3}
 H_{D,3}(A_n)=\log\!\left(8\binom{n+3}{3}\right)-\frac{S_3(n)}{8\binom{n+3}{3}}.
\end{equation}
\end{theorem}
\begin{proof}
Write $M=n+3$ and let $R$ be the number of runs of $x$. Put $d=M-R$ and let $b$ be its first bit. An alternating subsequence uses at most one position in each run. It starts in a run of index congruent to $1+b$ modulo two, and consecutive chosen run indices differ by an odd positive integer. A necessary and sufficient condition for a positive count is $R\ge n+b$. Consequently $d\le3-b$.

If $d=3$, then $R=n$ and only $b=0$ is possible. Every run is selected, and the multiplicity is the product of the run lengths. Distributing the three extra symbols gives one run of length four, one of length three and one of length two, or three runs of length two. The multiplicities and counts are respectively
\[
 (4,n),\quad(6,n(n-1)),\quad(8,\binom n3).
\]
If $d=2$, then $R=n+1$. For $b=0$ exactly the first $n$ runs are selected; for $b=1$ exactly the last $n$ are selected. For either initial bit, distributing two extra symbols gives multiplicity one once (both extras in the excluded run), multiplicity two $n$ times (one extra in the excluded run), multiplicity three $n$ times (both extras in one selected run), and multiplicity four $\binom n2$ times (extras in two selected runs).

If $d=1$, there is one run of length two among $R=n+2$ runs. For $b=1$, the selected runs are exactly $2,\ldots,n+1$: the multiplicity is one for the two boundary choices of the doubled run and two for its $n$ interior choices. For $b=0$, any run-selection pattern omits exactly one adjacent pair among $n+1$ pairs. Each boundary run is included in $n$ patterns and each interior run in $n-1$ patterns. Doubling a boundary run therefore changes the total from $n+1$ to $2n+1$, while doubling an interior run changes it to $2n$. Their respective counts are two and $n$.

Finally, $d=0$ gives the two alternating words of length $n+3$. Each has $n+1$ embeddings of $A_n$, as follows either by the gap argument in Theorem~\ref{thm:allI}, with pattern length $n$ and radius three, or directly by counting the run selections. Adding the disjoint cases gives the table and~\eqref{eq:SD3}. The entropy formula follows from Lemma~\ref{lem:entropy}.
\end{proof}

\section{Exact finite results and open comparisons}\label{sec:certificates}
\subsection{Enumeration and certificates}
No numerical logarithm is needed to compare entropies. If a histogram has $c_j$ distinct words of embedding multiplicity $j$, then $Q=\prod_j j^{j c_j}$ is an integer and $\Phi=\log Q$. At fixed received length and radius, normalization is constant, so integer ordering of $Q$ gives exact entropy ordering, including all ties.

For insertion, enumerate every received word $y\in\bits^n$ and every $k$-position deletion mask on it. Count how often each resulting input $x$ occurs, and then form its histogram. For deletion, enumerate every $x\in\bits^{n+k}$ and every $k$-position deletion mask on $x$. For each resulting received word $y$, accumulate the frequency $e(y,x)$ into $y$'s histogram. Each index set occurs exactly once in these loops. The former therefore computes exactly $e(x,y)$; the latter computes exactly $e(y,x)$, and every possible word is represented. This proves soundness and completeness of the finite enumeration.

The ancillary C++ enumerator performs these mask counts using integers. A Python checker verifies the normalization of every histogram, computes arbitrary-precision integer products, and compares every candidate against the minimum. It also checks the alternating histograms against the formulas in Theorem~\ref{thm:allI} and Theorem~\ref{thm:D3}. The saved outputs list all maximizers and their histograms, together with a strict runner-up whenever one exists.

\begin{proposition}\label{prop:finite}
The following finite statements hold by exhaustive exact comparison.
\begin{enumerate}
\item For three deletions and $1\le n\le12$, the maximizers are exactly $A_n$ and $\overline{A_n}$.
\item For three insertions and $7\le n\le18$, the maximizers are exactly $A_n$ and $\overline{A_n}$. The earlier cases are listed below.
\end{enumerate}
\begin{center}
\begin{tabular}{cll}
\toprule
$n$&three-insertion maximizing received words&minimum $Q$\\
\midrule
3&all words&1\\
4&the six words of weight two&$2^4$\\
5&$01010,01110,10001,10101$&$3^9$\\
6&$011001,100110$&$2^{20}3^6$\\
$7\text{--}18$&$A_n,\overline{A_n}$&$(n-2)^{(n-2)^2}$\\
\bottomrule
\end{tabular}
\end{center}
\end{proposition}
\begin{proof}
Apply the exhaustive algorithm above and compare~\eqref{eq:Q} for every received word. The ancillary results contain the exact comparisons for every length, including the $k=2$ initial cases used in Theorem~\ref{thm:two}. The three-deletion sweep covers $\sum_{n=1}^{12}2^n=8190$ received words; the three-insertion sweep covers $\sum_{n=3}^{18}2^n=524280$ received words. The programs use no symmetry reduction or floating-point filtering, so tied or exceptional words cannot be omitted by a preliminary numerical selection.
\end{proof}
For $n=6$, the three-insertion histogram of $011001$ has two occurrences of each multiplicity $1,2,3,4$. That of $A_6$ has four occurrences of each multiplicity $1,4$. Thus
\[
 Q(011001)=2^{20}3^6<2^{32}=Q(A_6),
\]
since $3^6<2^{12}$. Its entropy advantage is exactly $(12-6\log3)/20>0$. This is a concrete obstruction to extending the two-insertion maximizing statement without adjusting the initial lengths.

\subsection{Three-error conjectures}
\begin{conjecture}\label{conj:3}
For every $n\ge1$, the only maximizers of $H_{D,3}$ are $A_n$ and $\overline{A_n}$. For every $n\ge7$, the only maximizers of $H_{I,3}$ are $A_n$ and $\overline{A_n}$.
\end{conjecture}
The first part is the radius-three instance of the alternating-maximizer question discussed in~\cite{Atashpendar2019,Singhvi2025}. The second part incorporates the exact exceptions above. If these conjectures hold, their maximal values are~\eqref{eq:D3} and~\eqref{eq:I3}, respectively. Theorems~\ref{thm:allI} and~\ref{thm:D3} establish those alternating-word values at every length, independently of optimality. The remaining step is a strict comparison for arbitrary nonalternating words. The two-error proof suggests organizing this comparison by run excess and coincidences between masks, but at radius three these coincidences involve more than adjacent pairs.

\section*{Acknowledgement}
We thank Shubhransh Singhvi for his interest in a full account of the two-error techniques and for encouraging investigation of higher radii.

\section*{Data and code availability}
The enumeration code, exact outputs, reproduction instructions and Lean companion are supplied as arXiv ancillary files.
The maximizer characterizations and maximal-value formulas in Theorem~\ref{thm:two} are formalized in
\href{https://github.com/the-omega-institute/trureturing/blob/dev/D5/S3/Entropy/SinghviMaxInputEntropy.lean}{the corresponding Lean source}.
The higher-radius results and computations have the proofs and finite scope stated above.

\section*{Funding}
This research received no specific funding.

\section*{Competing interests}
The authors declare no competing interests.

\section*{Use of AI tools}
Haobo Ma and Wenlin Zhang used AI tools, namely ChatGPT Pro (GPT-6 Pro), Codex (gpt-6.1-sol) and Claude Opus 5.5 (via Claude Code), for literature search, computation and double-checking, including formal verification in Lean of Theorem~\ref{thm:two}. All proofs can be checked by hand, and the computations can be reproduced from the ancillary files. H. Ma and W. Zhang take full responsibility for their contributions.

\bibliographystyle{amsplain}
\begingroup
\small\raggedright
\bibliography{references}
\endgroup
\end{document}